\documentclass[twocolumn,aps,prd,linenumbers]{revtex4}	
\usepackage{graphics}
\usepackage[caption=false]{subfig}
\usepackage{float}
\usepackage{amsmath}
\usepackage[mathscr]{euscript}
\usepackage{acronym}
\floatstyle{boxed} 
\usepackage{mathtools}
\usepackage{multirow}
\def\thercsid{\relax}
\def\rcsid#1{\def\next##1#1{\def\thercsid{##1}}\next}
\rcsid$Id: QNM.tex,v  Doc No. $
\renewcommand{\today}{\number\day\space\ifcase\month\or
January\or February\or March\or April\or May\or June\or
July\or August\or September\or October\or November\or December\fi
\space\number\year}	

\newcommand{\Msun}{\ensuremath{\mathrm{M}_\odot}}

\usepackage{draftwatermark}

\SetWatermarkAngle{90}
\SetWatermarkLightness{1.0}
\SetWatermarkFontSize{0.5cm}
\SetWatermarkScale{1}
\begin{document}
\title{Corrected Version :  Spectroscopic analysis of stellar mass black-hole mergers in our local universe with ground-based gravitational wave detectors}

\author{Swetha Bhagwat, Duncan A. Brown, Stefan W. Ballmer$^{1}$}
\address{$^{1}$Syracuse University, Syracuse, NY 13244, USA}

\begin{abstract}

Motivated by the recent discoveries of binary black-hole mergers by the Advanced Laser Interferometer Gravitational-wave Observatory (Advanced LIGO), we investigate the prospects of ground-based detectors to perform a spectroscopic analysis of signals emitted during the ringdown of the final Kerr black-hole formed by a stellar mass binary black-hole merger. If we assume an optimistic rate of 240 Gpc$^{-3}$yr$^{-1}$, about 3 events per year can be measured by Advanced LIGO. Further, upgrades to the existing LIGO detectors will increase the odds of measuring multiple ringdown modes significantly. New ground-based facilities such as Einstein Telescope or Cosmic Explorer could measure multiple ringdown modes in about thousand events per year. We perform Monte-Carlo injections of $10^{6}$ binary black-hole mergers in a search volume defined by a sphere of radius 1500 Mpc centered at the detector, for various proposed ground-based detector models. We assume a uniform random distribution in component masses of the progenitor binaries, sky positions and orientations to investigate the fraction of the population that satisfy our criteria for detectability and resolvability of multiple ringdown modes. We investigate the detectability and resolvability of the sub-dominant modes $l=m=3$, $l=m=4$ and $l=2, m=1$. Our results indicate that the modes with $l=m=3$ and $l=2, m=1$ are the most promising candidates for sub-dominant mode measurability. We find that for stellar mass black-hole mergers, resolvability is not a limiting criteria for these modes. We emphasize that the measurability of the $l=2, m=1$ mode is not impeded by the resolvability criterion.

\end{abstract}

\maketitle
\section{\label{intro}Introduction}

The recent detection of gravitational waves from the coalescence of binary black-holes~\cite{PhysRevLett.116.061102,PhysRevLett.116.241103} stand as the first stringent test of the validity of the General Theory of Relativity in the regime of strong-field gravity~\cite{2013LRR....16....9Y,2016PhRvL.116v1101A,2016arXiv160308955Y}. We investigate whether detections of stellar-mass black-holes can be used to experimentally confirm some
fundamental predictions of this theory like the uniqueness
theorem and the no-hair theorem ~\cite{Misner:1974qy,PhysRevD.5.1239}. The no-hair theorem states that a space-time dictated by an isolated and stationary black-hole is fully characterized by just three parameters - the mass, the spin and the charge of the black-hole~\cite{PhysRevD.51.R6608,PhysRevLett.26.331}. Verifying the no-hair theorem would place strong constraints on possible alternative theories of gravitation~\cite{0264-9381-33-5-054001,0264-9381-32-21-214002}. In a binary black-hole system, the two black-holes orbit around each other eventually merging and settling down to a final stationary Kerr black-hole. This post-merger signal, often called ringdown, contains information about the final Kerr black-hole formed by the coalescence of of the progenitor black-holes \cite{0264-9381-21-4-003}, presenting us with an opportunity to verify the no-hair theorem. In light of these observations, efforts were made to study the ringdown signal. Although the features of the black-hole ringdown were discernible and had frequencies in a favorable regime of the detector's response, the signal-to-noise ratios (SNRs) of the signal in the two Laser Interferometer Gravitational-wave Observatory (LIGO) detectors were inadequate to perform a detailed ringdown analysis to draw firm conclusions about the final black-hole properties~\cite{2016PhRvL.116v1101A}. 

The ringdown signal seen by a distant observer during the coalescence of a binary black-hole system can be modeled as the gravitational waves arising from the perturbations, on the metric, associated with the final Kerr black-hole~\cite{PhysRevLett.72.3297}. At spatial asymptotic infinity, these perturbations on the Kerr background manifest themselves as superpositions of damped sinusoidal oscillating modes, known as quasi-normal-modes (QNMs)~\cite{4783,1973ApJ...185..635T,10.2307/78902,PhysRevD.40.3194,Nollert:1999ji,PhysRevD.60.022001}. Assuming the General Theory of Relativity is valid, the no-hair theorem necessitates that the spectrum of frequencies and the damping of these modes be dictated entirely by the mass and the spin of the final Kerr black-hole formed. Thus, a spectral analysis of the ringdown part of the signal not only helps us to understand the properties of the final black-hole formed, but also can serve as a test of the no-hair theorem.

We attempt to address the following three questions. What are the prospects of performing black-hole spectroscopy using future ground-based gravitational-wave detectors? Which of the modes contained in the ringdown are likely to be measurable? What is the frequency range that should be targeted to optimize sensitivity of ground-based detectors to test the no-hair theorem with the ringdown signal?

Our study concentrates on stellar mass black-hole mergers in our local universe. We focus our analysis on the measurability of the three largest sub-dominant modes: $l=m=3$, $l=m=4$ and $l=2, m=1$. We perform a Monte-Carlo injection of $10^{6}$ analytical post-merger gravitational wave signals, which are modeled as damped sinusoids with frequencies and damping times predicted by the linear perturbation theory on the Kerr background~\cite{Leaver285}. We do a mode-by-mode analysis; we consider each mode separately to assess its detectability and resolvability from the fundamental $l=m=2$ mode. We calculate the fraction of simulated signals that allow for measurability of at least one sub-dominant mode as well as the the dominant $l=m=2$ mode. We repeat this study with different proposed ground-based detectors - A+~\cite{2015PhRvD..91f2005M}, Einstein Telescope~\cite{0264-9381-28-9-094013} and Cosmic Explorer~\cite{Evans:2016dc}. A mode is considered detectable if its SNR is greater than 5 and resolvable if it satisfies the Rayleigh resolvability criterion~\cite{1453789,1038162} described in Section II. If a mode satisfies both of these conditions, we identify that mode as measurable. A signal with more than one measurable mode is spectroscopically valuable. Using the range of binary black-hole coalescence rates~\cite{LIGO:2016Rates} measured by Advanced LIGO~\cite{2015CQGra..32g4001L}, we estimate the number of spectroscopically valuable events per year.

Although it is improbable that we detect signals of spectroscopic value with Advanced LIGO, we estimate that hundreds of such signals will be detected by the future ground-based detectors such as Einstein Telescope and Cosmic Explorer every year. We deduce that the modes with $l=m=3$ and $l=2, m=1$ are the most promising candidates for sub-dominant mode measurability. Further, we find the measurability of the $l=2, m=1$ mode is not impeded by the resolvability criterion. We find that sub-dominant mode detectability is a sufficient condition to ensure measurability for all the modes considered in our study. We propose that a detector de-tuning around a frequency range of 300-500 Hz would be optimal for ringdown-oriented searches. 

Our work is complimentary to the recent work by Berti et al.~\cite{2016arXiv160509286B}. We perform a numerical Monte-Carlo simulation over sky positions and orientations and assume a uniform distribution of component masses of progenitor black-holes, while Berti et al. perform an approximate analytical angle-averaged analysis for different astrophysical black-hole population models. We have used a method of mode-by-mode matched filtering followed by a Fisher matrix analysis~\cite{van2004detection} to arrive at our results, in contrast to hypothesis testing and generalized likelihood used in~\cite{2016arXiv160509286B}. Another novel aspect of our work is that we include the $l=2, m=1$ sub-dominant mode. Although the two analyses differ in their methods, we agree on the result that a detector beyond Advanced LIGO is essential for spectroscopic analyses of black-hole mergers. 

The remainder of this paper is structured in the following way. In Section II we provide a detailed description of the analysis methods used in our study. Section III presents our results and highlights their implications to the broader theme of black-hole spectroscopy. We then conclude in Section IV on the prospects of stellar mass black-hole spectroscopy in our local universe with next-generation ground-based gravitational-wave detectors.

\section{Methods}
\label{sec:framework}

\begin{figure}
\includegraphics[width=\columnwidth]{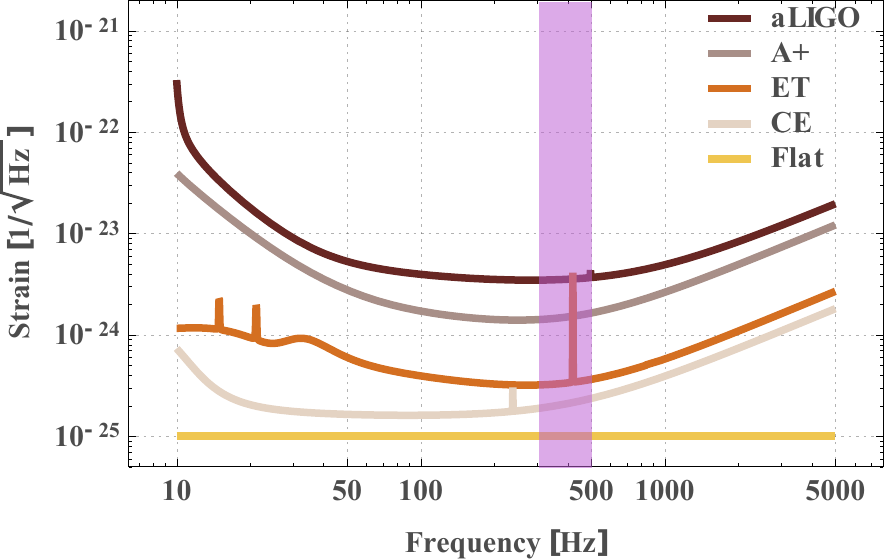}
\caption{\label{fig:DetectorCurve}The following are sensitivity models for each detector~\cite{Evans:2016dc} we consider in our study. The aLIGO curve corresponds to the design sensitivity of Advanced LIGO and the A+ curve to the proposed upgrade to the Advanced LIGO detectors. The Cosmic Explorer (CE) and the Einstein Telescope (ET) are two of the proposed next generation ground-based detectors. We also perform the analysis with a flat noise curve at a strain per $\sqrt[2]{\mathrm{Hz}}$ of $10^{-25}$, to infer some conclusions which are independent of the shape of the noise curve. The shaded region shows the frequency band that corresponds to optimal tuning of the detectors for ringdown searches. }
\end{figure}

We perform Monte-Carlo injections of ringdown-only gravitational-wave signals corresponding to stellar mass binary black-hole mergers in our local universe. Specifically, $10^{6}$ binary black-hole merger events are simulated uniformly in a volume defined by a sphere of radius 1500 Mpc around the the detector in question. Focusing our study on stellar mass mergers, we choose the component masses of the progenitor binary systems to be a uniform random distribution between 10~\Msun and 60~\Msun. Our study is limited to systems whose progenitor binaries are non-spinning, although we expect the qualitative results to hold for spinning cases. The sky positions and orientations of progenitor binaries with respect to the detector are also assumed to have a uniform random distribution. The analysis is performed independently for two future generation detectors - Einstein Telescope and Cosmic Explorer, a proposed upgrade to the current Advanced LIGO detector that we refer to as A+ and the design sensitivity of Advanced LIGO. 
Further, it is desirable to investigate which part of the detector's frequency range needs to be tuned for a ringdown-optimized search in addition to discerning which mode is more measurable, independent of the detector's sensitivity curve. For that reason, we also repeat the entire analysis on an unrealistic flat noise spectrum with a strain amplitude of $10^{-25}$ per $\sqrt[2]{\mathrm{Hz}}$ and present its results. For comparison, we display the relative sensitivities of all the detector curves used in our study in Figure~\ref{fig:DetectorCurve}.

We assume the binary black-hole ringdown signals observed by gravitational-wave detectors comprise of linear superpositions of a finite number of QNMs. Despite the mathematical issues, such as the incompleteness of QNMs~\cite{1999CMaPh.204..397B,lrr-1999-2}, it is known that for binary black-hole mergers this is a good approximate model ~\cite{2008GReGr..40.1705R, 2006PhRvD..74j4020B}. We test each mode independently for its measurability. Since we conduct a mode-by-mode analysis, we model the signal waveform as a single damped sinusoid of the following form:

\begin{align}
h_{lm}^{(+, \times)} (t) = \frac{M}{r}\left[A_{lm}^{(+,\times)} \mathrm{sin}(2 \pi f_{lm} t) e^{\frac{-t}{\tau_{lm}}} \mathcal{Y}_{lm}(\iota, \beta) \right].
\end{align}

Here $A_{lm}^{+, \times}$, $f_{lm}$ and $\tau_{lm}$ denote the amplitudes associated with the two polarizations, the central frequency and the damping-time, respectively, of the dominant overtone of $(l,m)$ modes in a black-hole ringdown. $(\iota, \beta)$ specify the orientation of the progenitor binary system in the sky. Further, we approximate the spheriodal harmonic function associated to each mode by spin-2 weighted spherical harmonics $\mathcal{Y}_{lm}(\iota, \beta)= \mathcal{Y}_{lm}(f_{lm};\iota, \beta)$, which is a good first order approximation for Kerr black-holes that are not extremaly spinning~\cite{2014PhRvD..90f4012B}.

\begin{figure}[h!]
\includegraphics[width=\columnwidth]{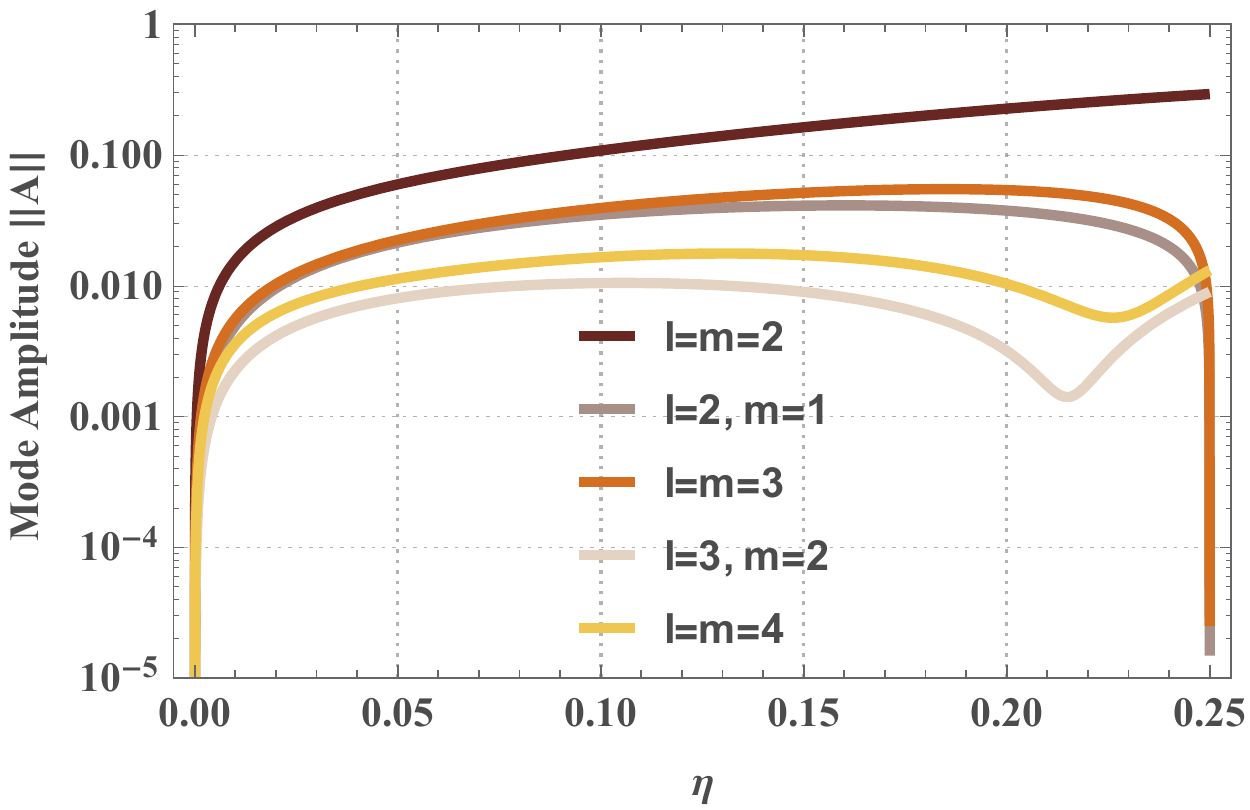}
\caption{\label{fig:ModeAmp}This figure presents the magnitude of mode amplitudes $||A||$ predicted by the fitting formulae given in~\cite{2014PhRvD..90l4032L} as a function of dimensionless symmetric-mass-ratio $\eta$. Comparing the amplitudes of different modes, we infer that the potential candidates for sub-dominant mode measurability correspond to $l=m=3$, $l=2, m=1$ and $l=m=4$.}
\end{figure}

We calibrate the central frequency and the decay time of each mode using the fitting functions presented in~\cite{PhysRevD.73.064030}. Ref.~\cite{2014PhRvD..90l4032L} presents mode amplitudes as functions of symmetric-mass-ratios $\eta$ of the progenitor binary system by fitting 68 numerical relativity waveforms corresponding to non-spinning black-hole binary systems. We have used the corrected formulae from the erratum ~\cite{Lionel:comm} for our analysis. The start-time of all modes are chosen to be 10 M after the occurrence of the peak in luminosity corresponding to the $l=m=2$ mode. We use these fitting formulae to determine the mode amplitudes $A_{lm}$ in our waveform model. Figure~\ref{fig:ModeAmp} presents the mode amplitudes of the sub-dominant modes. Dictated by the symmetry of the initial perturbation, the $l=m=2$ mode is the dominant mode in the ringdown of a Kerr black-hole formed during the merger of a binary black-hole system. Based on the sub-dominant mode amplitudes, we limit the scope of this study to $l=m=3$, $l=m=4$ and $l=2, m=1$ sub-dominant mode measurability. 

The signal $h(t)$ observed at a detector is then given as, 

\begin{align}
h(t)=F_{+}h_{+}(t) +F_{\times }h_{\times}(t),
\end{align}
where $F_{+, \times}$ are orientation-dependent detector pattern functions that project the signal on to the detector.

\begin{figure}[b!]
\includegraphics[width=\columnwidth]{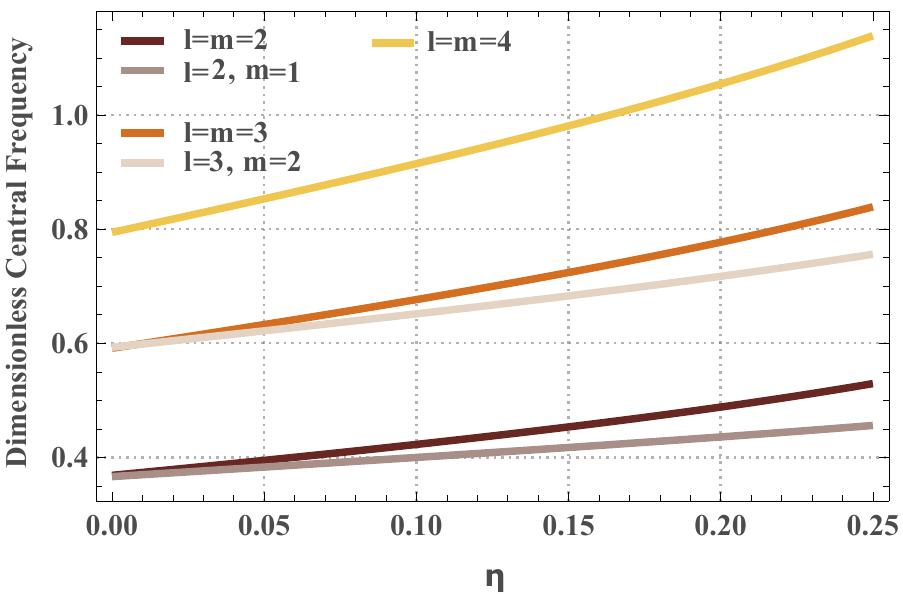}
\caption{\label{fig:DimFreqPlot}We show the dimensionless central frequency of QNMs as a function of symmetric-mass-ratio $\eta$ as predicted by~\cite{PhysRevD.73.064030}. Note that modes with different $l$ have central frequencies that are well separated. One could naively expect that resolving modes with the same $l$ could be challenging. However, for stellar mass black-hole mergers this is not the case.}
\end{figure}

\begin{figure}[hpt!]
\centering
\subfloat[Difference between central frequencies of $l=m=2$ and $l=2,m=1$ modes in Hz. ]{\label{figur:1}\includegraphics[width=68mm]{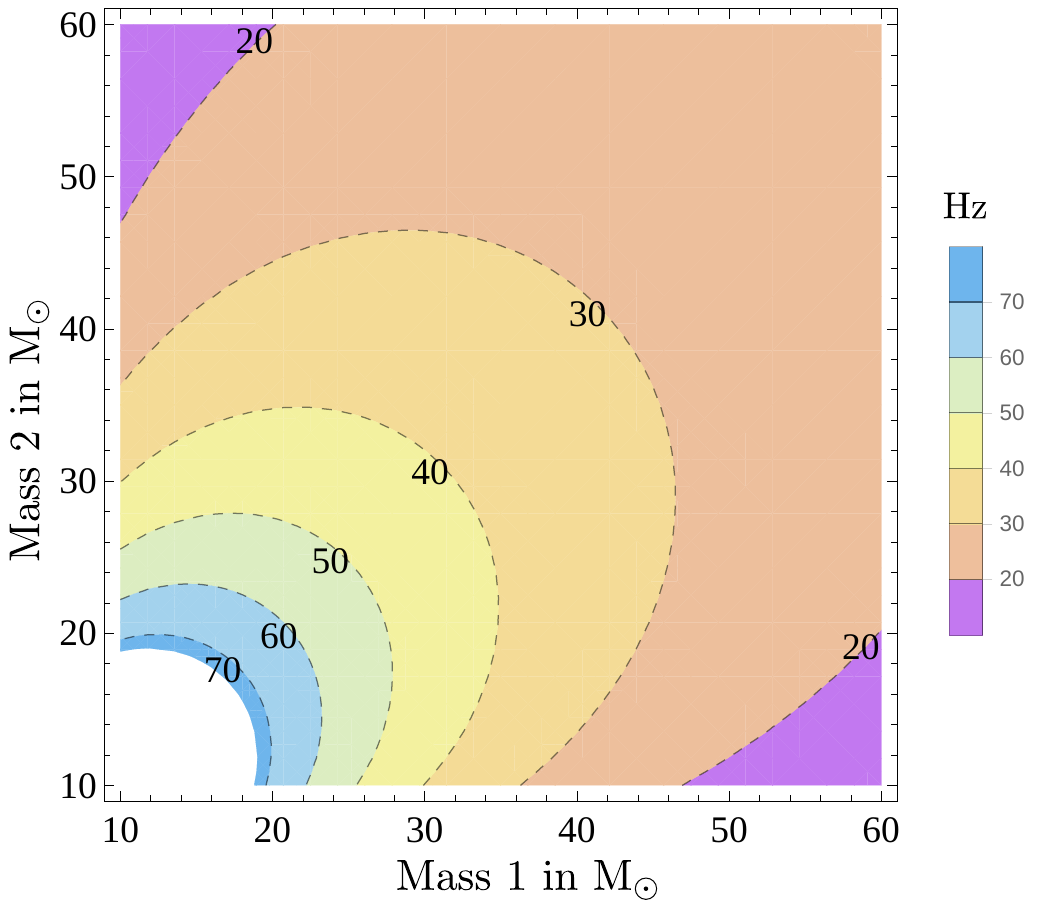}}
\\
\subfloat[Difference between central frequencies of $l=m=2$ and $l=m=3$ modes in Hz.]{\label{figur:2}\includegraphics[width=68mm]{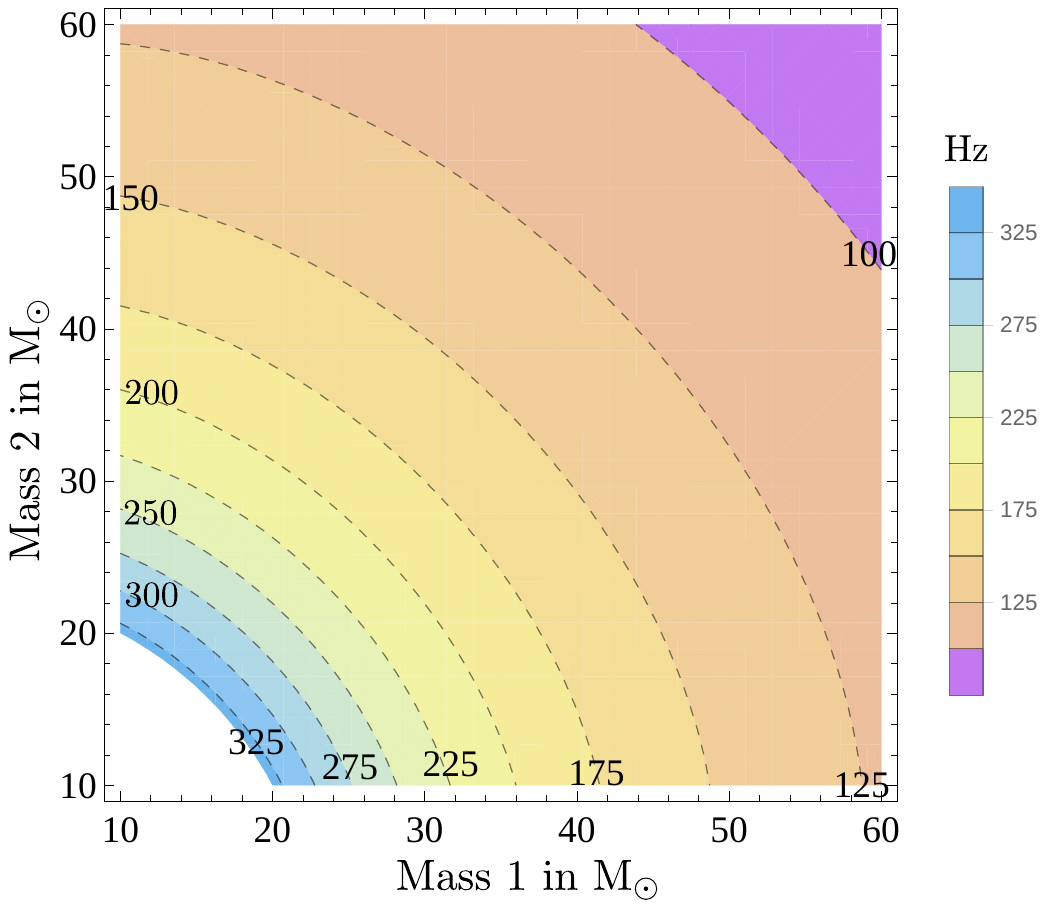}}
\\
\subfloat[Difference between central frequencies of $l=m=2$ and $l=m=4$ modes in Hz.]{\label{figur:3}\includegraphics[width=68mm]{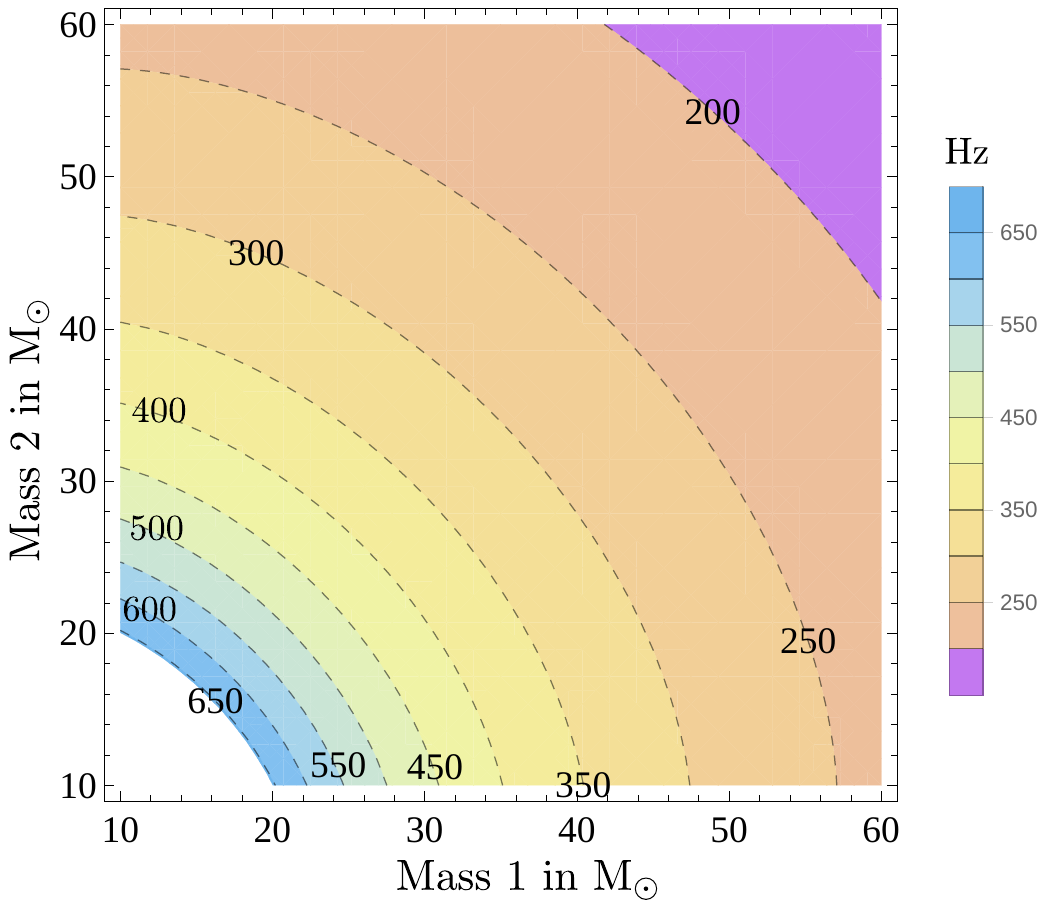}}

\caption{\label{fig:contourPlot}These contour plots show the differences in the central frequencies of the sub-dominant modes: $l=2,m=1$, $l=m=3$ and $l=m=4$, with the dominant mode. The color bar presents a measure of frequency difference in Hz. Notice that the central frequency of the $l=m=3$ and $l=m=4$ sub-dominant mode differs from the dominant mode by hundreds of Hz. It would be right to assume that resolvability of these modes is not challenging. However, it is very interesting to note that even for the $l=2, m=1$ sub-dominant mode the central frequency is separated by at least 20 Hz from the central frequency of the dominant mode. This is consistent with the fact that our results indicate that resolvability is not a limiting factor.}
\end{figure}

Expressing this in the Fourier domain, we obtain

\begin{gather}
\tilde{h_{+}}(f)=\frac{M}{r} A^{+}_{lm}\left[e^{\iota \phi^{+}_{lm}} \mathcal{Y}_{lm} b_{+} + e^{- \iota \phi^{+}_{lm}} \mathcal{Y^{*}}_{lm} b_{-} \right]
\\ \tilde{h_{\times}}(f)=-\iota \frac{M}{r} A^{\times}_{lm}\left[e^{\iota \phi^{\times}_{lm}} \mathcal{Y}_{lm} b_{+} - e^{- \iota \phi^{\times}_{lm}} \mathcal{Y^{*}}_{lm} b_{-} \right],
\end{gather}

where

\begin{gather}
b_{\pm}=\frac{2/\tau_{lm}}{(1/\tau_{lm})^{2} + {2 \pi (f \pm f_{lm})^{2}}}
\end{gather}

and $\phi^{+}_{lm}$ and $\phi^{\times}_{lm}$ are phase of arrival associated with $\tilde{h_{+}}(f)$ and $\tilde{h_{+}}(f)$ respectively. We follow ~\cite{2006PhRvD..73f4030B} in setting up the framework for our analysis. 

We use the standard expression for SNR $\rho$
\begin{align}
\rho^{2}=4 \int_{0}^{\infty} \frac{\tilde{h^{*}}(f')\tilde{h}(f')}{S_{h}(f')} df' = \langle h|h\rangle,
\end{align}
where $\tilde{h}(f)$ is the Fourier transform of the waveform and $S_{h}(f)$ is the power spectral density of the detector~\cite{PhysRevD.49.2658}. A mode is considered detectable if the single detector SNR of that mode exceeds a pre-defined threshold for detection. We choose $\rho \geq 5$ as our threshold and each mode is independently checked for this detectability criterion. Once a sub-dominant mode passes this criterion for detectability, we then proceed to check that its central frequency is resolvable from that of the dominant mode. 

We use an extension of Rayleigh criterion developed in~\cite{PhysRevD.73.064030,1453789,1038162} to establish the limits of resolvability. The Rayleigh criterion for diffraction states that to distinguish two points, the diffraction maxima of the second point should lie at least at the minima of the first point~\cite{van2004detection}. This translates to a condition that the peak of the estimators of QNM frequencies should be separated by at least the largest of their variances. If $\sigma_{f1}^2$ and $\sigma_{f2}^2$ are the variances of the maximum likelihood estimators of $f_{1}$ and $f_{2}$ associated with the modes under investigation, then the minimum criterion for resolvability is given by,
\begin{align}
 \frac{\textbf{max} [\sigma_{f_{1}},\sigma_{f_{2}}]}{|f_{1}-f_{2}|}=1.
\end{align}

In the scheme of Fisher information theory, the spread $\sigma_{f_{i}}$ in the estimate of the frequency $f_{i}$ is given by

\begin{align}
\sigma_{f_{i}}^{2}=\Gamma_{f_{i}f_{i}}^{-1},
\end{align}
where $\Gamma$ is the Fisher matrix~\cite{PhysRevD.46.5236}. To compute the Fisher matrix, we parametrize the waveform by the mode amplitude, frequency, quality factor, arrival time and phase. The likelihood function has peaks around the central frequency of each of the QNMs. We perform a Fisher matrix analysis around each of these mode frequencies to determine the spread in the estimate of the central frequency of the modes. Then it follows that the critical SNR $\rho_{c}$ that sets the resolvability limit of these modes is given by,
\begin{gather}
\rho_{c}=\frac{\textbf{max} [\rho \sigma_{f_{1}},\rho \sigma_{f_{2}}]}{|f_{1}-f_{2}|}.
\end{gather}
A dimensionless ratio $\mathcal{R}$ determines the resolvability of QNMs:

\begin{gather}
\mathcal{R}=\frac{\rho}{\rho_{c}}=\frac{|f_{1}-f_{2}|}{\textbf{max} [\sigma_{f_{1}}, \sigma_{f_{2}}]}.
\end{gather}
When $\mathcal{R}$ is greater than 1, the central frequency of the sub-dominant mode in the signal can be successfully resolved from the dominant mode. 

Having established our criteria of detectability and resolvability, we perform a mode-by-mode analysis on each of the injected signals with the detector curves depicted in Figure~\ref{fig:DetectorCurve}. Equations (6) and (10) are evaluated numerically and for each mode we test if $\rho_{lm} > 5$ and $\mathcal{R} > 1$ to determined their measurability. We then categorize the signals based on their measurability.

\begin{figure}[h!]
\centering

\subfloat[Central frequencies of modes $l=m=3$ and $l=m=2$ in Hz.]{\label{figur:1}\includegraphics[width=75mm]{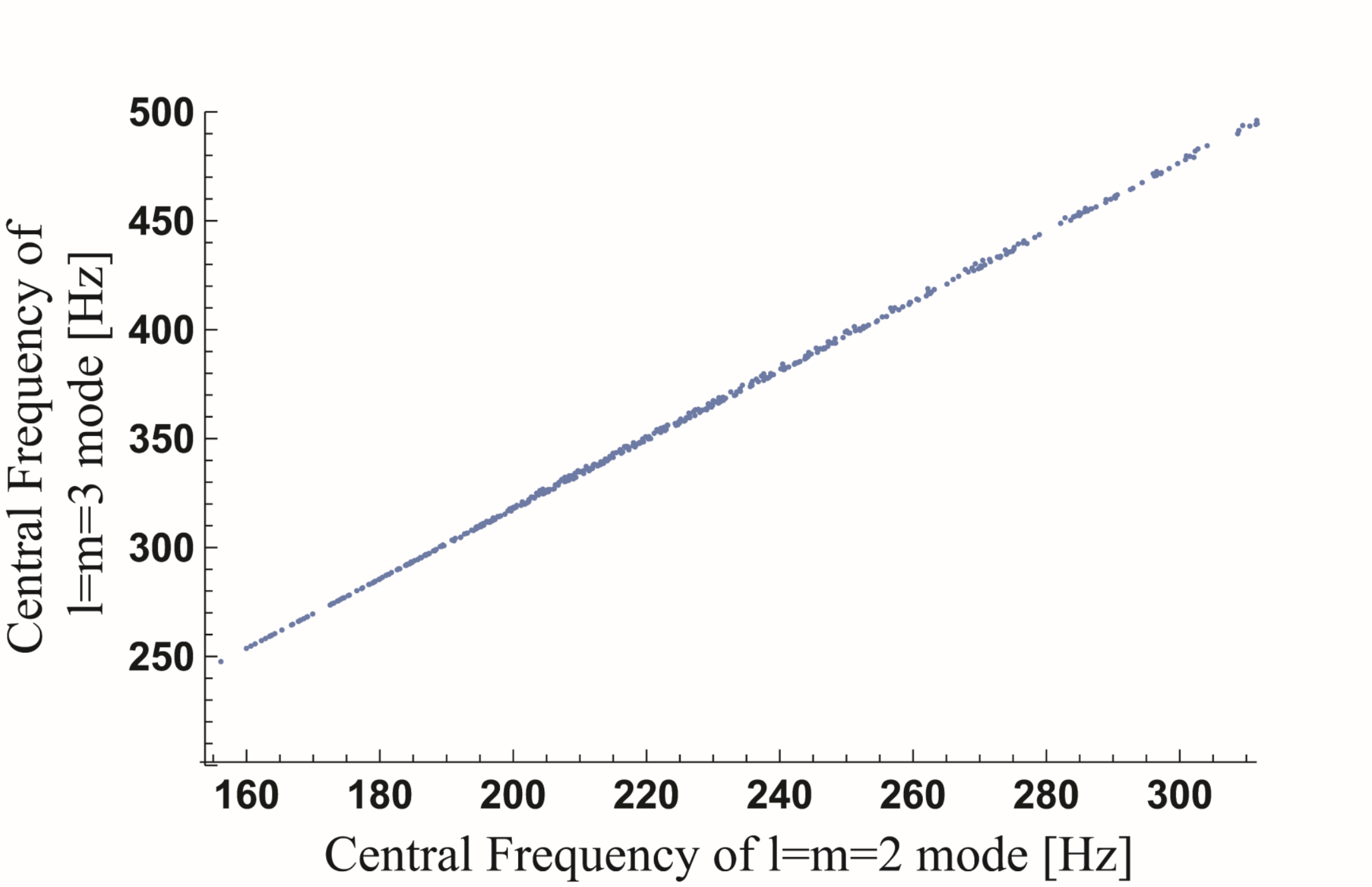}}
\\
\subfloat[Central frequencies of modes $l=m=4$ and $l=m=2$ in Hz.]{\label{figur:2}\includegraphics[width=75mm]{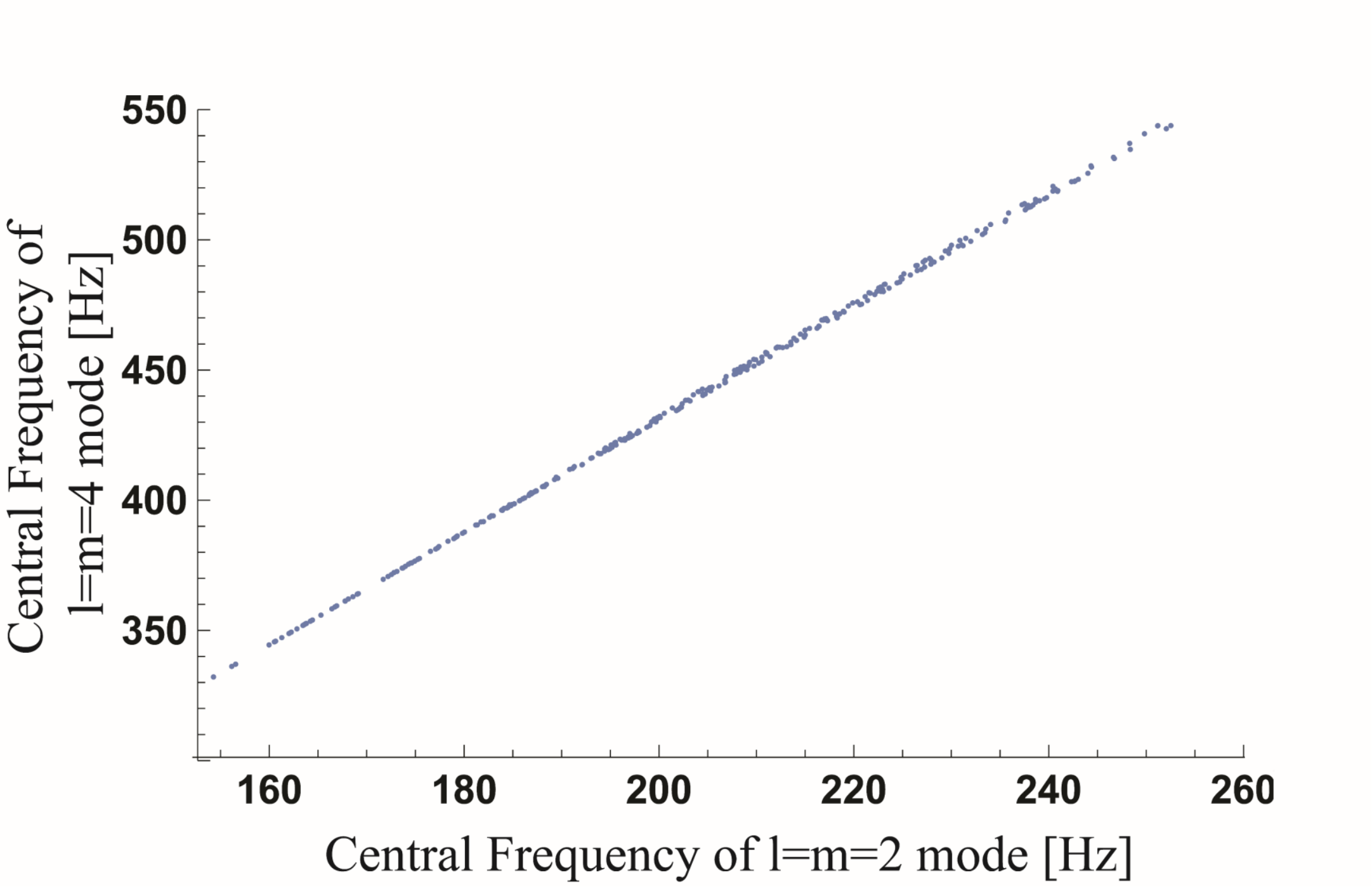}}
\\
\centering
\subfloat[Central frequencies of modes $l=2, m=1$ and $l=m=2$ in Hz.]{\label{figur:2}\includegraphics[width=75mm]{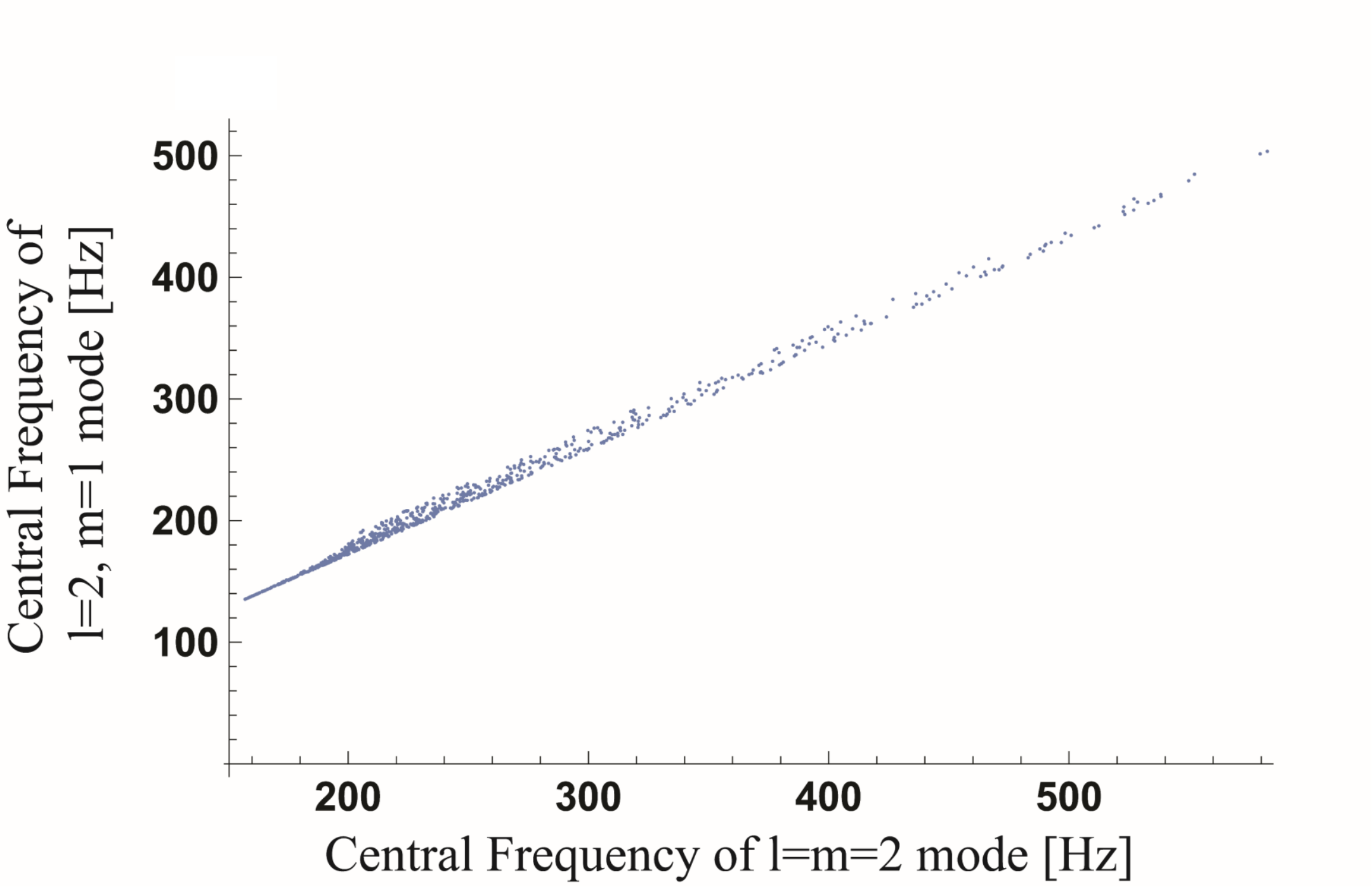}}

\caption{\label{fig:freqInHz}Scatter plots of all points that allow for measurability of sub-dominant modes in our analysis using a flat detector sensitivity curve at a strain of $10^{-25}$$\sqrt[2]{\mathrm{Hz}}^{-1}$. The x and y axes of these plots correspond to the central frequencies of $l=m=2$ and the measurable sub-dominant modes in Hz respectively. From these plots, we can infer that if one were to perform detector de-tuning optimized towards a spectroscopic analysis of stellar mass black-holes, a frequency band around 300 Hz to 500 Hz would be the best choice for narrow banding.}
\end{figure}

\section{Results and Implications}
\label{sec:results}

We find that we are able to measure sub-dominant modes during the ringdown of stellar mass Kerr black-holes with the proposed designs for future ground-based gravitational wave detectors. The results are summarized in Tables 1-4. With detectors like Cosmic Explorer and Einstein Telescope, we find that approximately 20-30\% of the total detected stellar mass black-hole mergers will be spectroscopically valuable. Our results also indicate that the design sensitivity of Advanced LIGO might detect signals that would allow for multi-mode measurements if the optimistic rate predictions hold. However, implementing A+ to the detector would increase our odds of sub-dominant mode measurability to a little less than 3\% of the total detected black-hole mergers.

The astrophysical rates of stellar mass black-hole mergers have significant uncertainty and hence, in Table 3, we tabulate both the optimistic and pessimistic rates of events that would allow for ringdown spectroscopy of the final Kerr black-hole using various proposed ground-based detectors. With an optimistic rate of 240 Gpc$^{-3}$yr$^{-1}$ merger events~\cite{LIGO:2016Rates}, we expect about a thousand events per year will be spectroscopically valuable with Einstein Telescope and Cosmic Explorer. It is further encouraging to notice that, with the implementation of the A+ upgrade to the current detectors, an optimistic rate would indicate that about an order of 50 spectroscopically valuable events will be detected every year. Further, even a pessimistic rate of only 13 Gpc$^{-3}$yr$^{-1}$ binary black-hole mergers, leads us to estimate about 40-60 events that allows for multi-mode measurements using Einstein Telescope and Cosmic Explorer.

From our analysis using a flat detector curve depicted in Figure~\ref{fig:DetectorCurve} we infer, independent of proposed-detector sensitivities, that the $l=m=3$ sub-dominant mode has the most measurability, closely followed by the sub-dominant mode with $l=2, m=1$. An optimistic rate of 240 Gpc$^{-3}$yr$^{-1}$ merger events suggests that nearly 650-1000 events would allow for measurability of the $l=m=3$ sub-dominant mode and about a 100-250 would allow for measurability of the $l=m=4$ mode each year with Cosmic Explorer and Einstein Telescope. Furthermore, analyzing the mode $l=2, m=1$, we find that its measurability with Cosmic Explorer is about 1000 events per year and that with Einstein Telescope is about a 500 events per year. A few of the existing literature~\cite{2004CQGra..21..787D,2016arXiv160509286B} have not considered the sub-dominant mode corresponding to $l=2, m=1$ in their studies. Our study highlights that for the detection of stellar mass black-hole mergers with Cosmic Explorer, the $l=2, m=1$ is the most promising mode. From Figure~\ref{fig:ModeAmp}, we can see that the $l=2, m=1$ sub-dominant mode has a slightly smaller mode amplitude compared to the $l=m=3$ mode. However it should also be noted that it is the least damped sub-dominant mode. Thus, for noise curves such as that of Cosmic Explorer, where the detector has a favorable sensitivity in lower frequencies, the odds of measuring the $l=2, m=1$ sub-dominant mode is markedly elevated. 

In contrast to the naive expectation formed by looking at Figure~\ref{fig:DimFreqPlot}, we find that the frequency of the sub-dominant mode corresponding to $l=2, m=1$ is well separated from the central frequency of the dominant mode for the case of stellar mass black-hole mergers. For all the sub-dominant modes, including $l=2, m=1$ we notice that detectability is the primary condition that limits mode measurability and that only very few signals fail measurability due to the resolvability criterion. Figure~\ref{fig:contourPlot} shows that even for $l=2, m=1$ mode, the central frequency of the dominant and sub-dominant mode differ by at least 20 Hz for all cases considered in our study. Thus, resolvability does not seem to crucially effect measuarbility of the modes. This result might be advantageous while developing new data-analysis techniques to measure sub-dominant modes for stellar mass black-hole mergers because it indicates that checking detectability is sufficient and removes an additional layer of complexity of having to check mode resolvability. However, resolvability could indeed become a potential challenge if one were to deal with super-massive black-holes targeted by planned space-based detectors. In such cases a more carefully designed data-analysis technique needs to be developed.

Finally, we address the question of which frequency band should be targeted for a ringdown oriented detector de-tuning. For a spectroscopic analysis of black-hole ringdowns, our focus should be on measuring the sub-dominant modes because their single-mode SNRs are generally much smaller than the dominant mode. The scatter plots in Figure~\ref{fig:freqInHz} capture the information of mode frequencies corresponding to the population of signals that passed our measurability criterion. Again, this plot is made using a flat sensitivity curve to arrive at a conclusion that is independent of the shape of the detector noise curve. Looking at the central frequencies of sub-dominant modes $l=m=3$ and $l=m=4$ of signals that passed out measurability criterion in Figure~\ref{fig:DimFreqPlot} and Table 2, we propose that an increase in sensitivity around 300 Hz and 500 Hz would enhance the measurability of both $l=m=3$ and $l=m=4$. Measurability of $l=2, m=1$ sub-dominant mode however would benefit from detector de-tuning around 150-300 Hz. Considering that the joint measurability of sub-dominant modes $l=m=3$ and $l=m=4$ seems more promising, it can be inferred that a frequency band between 300 Hz and 500 Hz is the best target for detector tuning optimized for spectroscopic analysis of stellar mass black-holes. This result for frequency tuning relies on the assumption that the the initial black holes are uniformly distributed in the mass range of 10 to 60 $\Msun$. Using a different astrophysical source distribution will lead to different optimal frequency bands, since the distribution of QNM frequencies depend on the masses of progenitors. Our method can be used to compute this frequency tuning for other mass distributions.

\section{Conclusion}
\label{sec:conc}

 In this paper we have investigated the prospects of our ability to perform black-hole spectroscopy using the current and future ground-based gravitational-wave detectors. We find that with a realistic rate of binary black-hole mergers, one could expect to detect several tens of spectroscopically valuable signals with future ground-based detectors like Einstein Telescope and Cosmic Explorer. Although Advanced LIGO might detect signals that would allow for multi-mode measurements only if the optimistic rates hold, implementing A+ upgrade increases our odds of detecting such signals. From the results of this study, we also conclude that sub-dominant modes corresponding to $l=m=3$ and $l=2, m=1$ offer the most measurability. We emphasize that resolvability is not a limiting factor for stellar mass black-hole mergers for all the modes we have considered in our study. Further, we propose that a detector de-tuning around a frequency band between 300 Hz and 500 Hz is optimal for a ringdown-oriented search. 

In this study we have used the choice made in~\cite{2014PhRvD..90l4032L} that all modes of ringdown begin 10 M after the peak of luminosity corresponding to the $l=m=2$ mode. This choice was motivated by the work pioneered in ~\cite{2012PhRvL.109n1102K}. Although there is no absolute framework to choose the start time of the ringdown, this is a conservative choice. Even with this conservative choice, we find an encouraging rate of detectable spectroscopically valuable signals using the future ground-based detectors. We intend to explore alternative choices, such as in~\cite{2007PhRvD..76f4034B} in a future study and we expect this will improve the chances of measuring sub-dominant modes significantly. Further, this study is done in the scheme of the Fisher information theory. Future work will follow this study with a full Bayesian parameter estimation like that in ~\cite{PhysRevD.85.124056} and a comparison of the results.


\begin{table*}[h!]
\centering
\begin{tabular}{||c c c c c c c||} 
 \hline
 Detector Curve & Set 1 & Set 2 & Set 3 & Set 4 & Set 5 & Set 6 \\ [0.5ex] 
 \hline\hline
 Advanced LIGO & 0.83 & 0.16 & 0.0008 &2 $\times 10^{-5}$ &1.6 $\times 10^{-5}$ &0\\ 
 A+ & 0.6 & 0.37 & 0.01 & 0.0004 &0.0001 &0\\
 Einstein Telescope & 0.32 & 0.47 & 0.18 &0.03 & 0.01 & 2 $\times 10^{-4}$\\
 Cosmic Explorer & 0.3 & 0.44 & 0.3 &0.08 & 0.04 &5 $\times 10^{-6}$\\
 A flat noise curve at $10^{-25}$ $\sqrt[2]{\mathrm{Hz}}^{-1}$ & 0.17 & 0.35 & 0.41 & 0.06 & 0.18 &0.001 \\ [1ex] 
 \hline
\end{tabular}
\caption{The above table shows the results we obtain from a Monte-Carlo simulation of $10^{6}$ stellar mass binary black-hole mergers uniformly distributed in component mass, orientation and in volume defined by a sphere of radius 1500 Mpc. We categorize each event into one of the set defined below and tabulate the fraction of signals that fall into each set. Set 1: $l=m=2$ mode could not be detected, Set 2: $l=m=2$ could be detected but no other sub-dominant mode could be detected, Set 3: $l=m=3$ sub-dominant mode can be measured, Set 4: $l=m=4$ sub-dominant mode can be measured, Set 5: Both $l=m=3$ and $l=m=4$ sub-dominant modes can be measured, Set 6: failed measurability of sub-dominant mode due to resolvability criterion.}
\label{table:1}
\end{table*}



\begin{table*}[h!]
\centering
\begin{tabular}{||c c c c c ||} 
 \hline
 Detector Curve & Set 1 & Set 2 & Set 3 & Set 4 \\ [0.5ex] 
 \hline\hline
 Advanced LIGO & 0.84 & 0.15& 1.8$\times 10^{-4}$ & 0 \\
 A+ & 0.59 & 0.4& 0.004 & 0 \\
 Einstein Telescope & 0.31 & 0.52 & 0.16 &0 \\
 Cosmic Explorer & 0.24 & 0.41 & 0.34 &0 \\
 A flat noise curve at $10^{-25}$ $\sqrt[2]{\mathrm{Hz}}^{-1}$ & 0.14 & 0.33 & 0.52 &0 \\
 [1ex] 
 \hline
\end{tabular}
\caption{The above table has information similar to Table 1 but with sets defined differently. Here, Set 1: $l=m=2$ mode could not be detected, Set 2: $l=m=2$ could be detected but $l=2, m=1$ sub-dominant mode could be detected, Set 3: $l=2, m=1$ sub-dominant mode is both detected and resolved, Set 4: $l=2, m=1$ sub-dominant mode is detected but not resolved . Here again, we tabulate the number of events out of $10^{6}$ Monte-Carlo simulated binary black-hole mergers that fall in each of these sets.}
\label{table:2}
\end{table*}

\begin{table*}[hbp!]
\centering
\begin{tabular}{||c c c ||} 
 \hline
 Detector Curve & Optimistic Rate & Pessimistic Rate \\ [0.5ex] 
 \hline\hline
 Advanced LIGO & 2.9/yr & 0.2/yr \\ 
 A+ & 47.3/yr & 2.6/yr \\
 Einstein Telescope & 694.3/yr & 37.6/yr \\
 Cosmic Explorer & 1130/yr & 61.2/yr \\
 A flat noise curve at $10^{-25}$ $\sqrt[2]{\mathrm{Hz}}^{-1}$ & 1570/yr & 85 /yr \\ [1ex] 
 \hline
\end{tabular}
\caption {Using our results in Table 1 and the optimistic (pessimistic) rates of binary black-hole mergers, predicted based on the recent discoveries of binary black-hole mergers~\cite{LIGO:2016Rates}, at $240$ Gpc$^{-3}$ yr$^{-1}$ ($13$ Gpc$^{-3}$ yr$^{-1}$ ), we present the rate of events that would allow measurability of $l=m=3$ or $l=m=4$ sub-dominant mode with current and future ground-based detectors. We present this combined ($l=m=3$ or $l=m=4$) rate, because de-tuning the detector around the frequency band $300-500$ Hz for a ringdown oriented search benefits both of these modes. }
\label{table:2}
\end{table*}




%

\begin{table*}
\centering
\subfloat 
\centering
\begin{tabular}{||c c c ||} 
 \hline
 For $l=m=3$& Optimistic Rate & Pessimistic Rate \\ [0.5ex] 
 \hline\hline
 Advanced LIGO & 2.9/yr & 0.2/yr \\ 
 A+ & 46.4/yr & 2.5/yr \\
 Einstein Telescope & 645.4/yr & 34.9/yr \\
 Cosmic Explorer & 1024.3/yr & 55.5/yr \\
 A flat noise curve at $10^{-25}$ $\sqrt[2]{\mathrm{Hz}}^{-1}$ & 1409.2/yr & 76.3/yr \\ [1ex] 
 \hline
\end{tabular}

\subfloat 
\centering
\begin{tabular}{||c c c ||} 
 \hline
 For $l=m=4$& Optimistic Rate & Pessimistic Rate \\ [0.5ex] 
 \hline\hline
 Advanced LIGO & 0.08/yr & 0.004/yr \\ 
 A+ & 1.4/yr & 0.08/yr \\
 Einstein Telescope & 96.7/yr & 5.2/yr \\
 Cosmic Explorer & 263.1/yr & 14.2/yr \\
 A flat noise curve at $10^{-25}$ $\sqrt[2]{\mathrm{Hz}}^{-1}$ & 605.7/yr & 32.8/yr \\ [1ex] 
 \hline
\end{tabular}

\centering
\begin{tabular}{||c c c ||} 
 \hline
For $l=2, m=1$ & Optimistic Rate & Pessimistic Rate \\ [0.5ex] 
 \hline\hline
 Advanced LIGO & 0.6/yr & 0.03/yr \\ 
A+ & 13.4/yr & 0.7/yr \\
 Einstein Telescope & 545.6/yr & 29.6/yr \\
 Cosmic Explorer & 1162.7/yr & 63/yr \\
 A flat noise curve at $10^{-25}$ $\sqrt[2]{\mathrm{Hz}}^{-1}$ & 1772.3/yr & 96/yr \\ [1ex] 
 \hline
\end{tabular}

\caption{Using our results in Table 2 and the optimistic (pessimistic) rate of binary black-hole mergers, predicted based on the recent discoveries of binary black-hole mergers~\cite{LIGO:2016Rates}, at $240$ Gpc$^{-3}$ yr$^{-1}$ ($13$ Gpc$^{-3}$ yr$^{-1}$ ), we present the rate of events that would allow measurability of single sub-dominant modes.}
\end{table*}

\acknowledgements
DAB and SB acknowledge support from NSF awards PHY-1404395 and AST-1333142. SWB acknowledges support from NSF award PHY-1352511. SB would like to thank Laura Nuttall, Ben Lackey, Lionel London and Emanuele Berti for providing useful suggestions.

\bibliography{Biblo}

\begin{thebibliography}{43}
\expandafter\ifx\csname natexlab\endcsname\relax\def\natexlab#1{#1}\fi
\expandafter\ifx\csname bibnamefont\endcsname\relax
  \def\bibnamefont#1{#1}\fi
\expandafter\ifx\csname bibfnamefont\endcsname\relax
  \def\bibfnamefont#1{#1}\fi
\expandafter\ifx\csname citenamefont\endcsname\relax
  \def\citenamefont#1{#1}\fi
\expandafter\ifx\csname url\endcsname\relax
  \def\url#1{\texttt{#1}}\fi
\expandafter\ifx\csname urlprefix\endcsname\relax\def\urlprefix{URL }\fi
\providecommand{\bibinfo}[2]{#2}
\providecommand{\eprint}[2][]{\url{#2}}

\bibitem[{\citenamefont{Abbott et~al.}(2016{\natexlab{a}})\citenamefont{Abbott,
  Abbott, Abbott, Abernathy, Acernese, Ackley, Adams, Adams, Addesso, Adhikari
  et~al.}}]{PhysRevLett.116.061102}
\bibinfo{author}{\bibfnamefont{B.~P.} \bibnamefont{Abbott}},
  \bibinfo{author}{\bibfnamefont{R.}~\bibnamefont{Abbott}},
  \bibinfo{author}{\bibfnamefont{T.~D.} \bibnamefont{Abbott}},
  \bibinfo{author}{\bibfnamefont{M.~R.} \bibnamefont{Abernathy}},
  \bibinfo{author}{\bibfnamefont{F.}~\bibnamefont{Acernese}},
  \bibinfo{author}{\bibfnamefont{K.}~\bibnamefont{Ackley}},
  \bibinfo{author}{\bibfnamefont{C.}~\bibnamefont{Adams}},
  \bibinfo{author}{\bibfnamefont{T.}~\bibnamefont{Adams}},
  \bibinfo{author}{\bibfnamefont{P.}~\bibnamefont{Addesso}},
  \bibinfo{author}{\bibfnamefont{R.~X.} \bibnamefont{Adhikari}},
  \bibnamefont{et~al.} (\bibinfo{collaboration}{LIGO Scientific Collaboration
  and Virgo Collaboration}), \bibinfo{journal}{Phys. Rev. Lett.}
  \textbf{\bibinfo{volume}{116}}, \bibinfo{pages}{061102}
  (\bibinfo{year}{2016}{\natexlab{a}}),
  \urlprefix\url{http://link.aps.org/doi/10.1103/PhysRevLett.116.061102}.

\bibitem[{\citenamefont{Abbott et~al.}(2016{\natexlab{b}})\citenamefont{Abbott,
  Abbott, Abbott, Abernathy, Acernese, Ackley, Adams, Adams, Addesso, Adhikari
  et~al.}}]{PhysRevLett.116.241103}
\bibinfo{author}{\bibfnamefont{B.~P.} \bibnamefont{Abbott}},
  \bibinfo{author}{\bibfnamefont{R.}~\bibnamefont{Abbott}},
  \bibinfo{author}{\bibfnamefont{T.~D.} \bibnamefont{Abbott}},
  \bibinfo{author}{\bibfnamefont{M.~R.} \bibnamefont{Abernathy}},
  \bibinfo{author}{\bibfnamefont{F.}~\bibnamefont{Acernese}},
  \bibinfo{author}{\bibfnamefont{K.}~\bibnamefont{Ackley}},
  \bibinfo{author}{\bibfnamefont{C.}~\bibnamefont{Adams}},
  \bibinfo{author}{\bibfnamefont{T.}~\bibnamefont{Adams}},
  \bibinfo{author}{\bibfnamefont{P.}~\bibnamefont{Addesso}},
  \bibinfo{author}{\bibfnamefont{R.~X.} \bibnamefont{Adhikari}},
  \bibnamefont{et~al.} (\bibinfo{collaboration}{LIGO Scientific Collaboration
  and Virgo Collaboration}), \bibinfo{journal}{Phys. Rev. Lett.}
  \textbf{\bibinfo{volume}{116}}, \bibinfo{pages}{241103}
  (\bibinfo{year}{2016}{\natexlab{b}}),
  \urlprefix\url{http://link.aps.org/doi/10.1103/PhysRevLett.116.241103}.

\bibitem[{\citenamefont{{Yunes} and {Siemens}}(2013)}]{2013LRR....16....9Y}
\bibinfo{author}{\bibfnamefont{N.}~\bibnamefont{{Yunes}}} \bibnamefont{and}
  \bibinfo{author}{\bibfnamefont{X.}~\bibnamefont{{Siemens}}},
  \bibinfo{journal}{Living Reviews in Relativity} \textbf{\bibinfo{volume}{16}}
  (\bibinfo{year}{2013}), \eprint{1304.3473}.

\bibitem[{\citenamefont{{Abbott} et~al.}(2016)\citenamefont{{Abbott}, {Abbott},
  {Abbott}, {Abernathy}, {Acernese}, {Ackley}, {Adams}, {Adams}, {Addesso},
  {Adhikari} et~al.}}]{2016PhRvL.116v1101A}
\bibinfo{author}{\bibfnamefont{B.~P.} \bibnamefont{{Abbott}}},
  \bibinfo{author}{\bibfnamefont{R.}~\bibnamefont{{Abbott}}},
  \bibinfo{author}{\bibfnamefont{T.~D.} \bibnamefont{{Abbott}}},
  \bibinfo{author}{\bibfnamefont{M.~R.} \bibnamefont{{Abernathy}}},
  \bibinfo{author}{\bibfnamefont{F.}~\bibnamefont{{Acernese}}},
  \bibinfo{author}{\bibfnamefont{K.}~\bibnamefont{{Ackley}}},
  \bibinfo{author}{\bibfnamefont{C.}~\bibnamefont{{Adams}}},
  \bibinfo{author}{\bibfnamefont{T.}~\bibnamefont{{Adams}}},
  \bibinfo{author}{\bibfnamefont{P.}~\bibnamefont{{Addesso}}},
  \bibinfo{author}{\bibfnamefont{R.~X.} \bibnamefont{{Adhikari}}},
  \bibnamefont{et~al.}, \bibinfo{journal}{Physical Review Letters}
  \textbf{\bibinfo{volume}{116}}, \bibinfo{eid}{221101} (\bibinfo{year}{2016}),
  \eprint{1602.03841}.

\bibitem[{\citenamefont{{Yunes} et~al.}(2016)\citenamefont{{Yunes}, {Yagi}, and
  {Pretorius}}}]{2016arXiv160308955Y}
\bibinfo{author}{\bibfnamefont{N.}~\bibnamefont{{Yunes}}},
  \bibinfo{author}{\bibfnamefont{K.}~\bibnamefont{{Yagi}}}, \bibnamefont{and}
  \bibinfo{author}{\bibfnamefont{F.}~\bibnamefont{{Pretorius}}},
  \bibinfo{journal}{ArXiv e-prints}  (\bibinfo{year}{2016}),
  \eprint{1603.08955}.

\bibitem[{\citenamefont{Misner et~al.}(1973)\citenamefont{Misner, Thorne, and
  Wheeler}}]{Misner:1974qy}
\bibinfo{author}{\bibfnamefont{C.}~\bibnamefont{Misner}},
  \bibinfo{author}{\bibfnamefont{K.}~\bibnamefont{Thorne}}, \bibnamefont{and}
  \bibinfo{author}{\bibfnamefont{J.}~\bibnamefont{Wheeler}},
  \emph{\bibinfo{title}{Gravitation}} (\bibinfo{publisher}{W.H. Freeman and
  Company}, \bibinfo{year}{1973}).

\bibitem[{\citenamefont{Bekenstein}(1972)}]{PhysRevD.5.1239}
\bibinfo{author}{\bibfnamefont{J.~D.} \bibnamefont{Bekenstein}},
  \bibinfo{journal}{Phys. Rev. D} \textbf{\bibinfo{volume}{5}},
  \bibinfo{pages}{1239} (\bibinfo{year}{1972}),
  \urlprefix\url{http://link.aps.org/doi/10.1103/PhysRevD.5.1239}.

\bibitem[{\citenamefont{Bekenstein}(1995)}]{PhysRevD.51.R6608}
\bibinfo{author}{\bibfnamefont{J.~D.} \bibnamefont{Bekenstein}},
  \bibinfo{journal}{Phys. Rev. D} \textbf{\bibinfo{volume}{51}},
  \bibinfo{pages}{R6608} (\bibinfo{year}{1995}),
  \urlprefix\url{http://link.aps.org/doi/10.1103/PhysRevD.51.R6608}.

\bibitem[{\citenamefont{Carter}(1971)}]{PhysRevLett.26.331}
\bibinfo{author}{\bibfnamefont{B.}~\bibnamefont{Carter}},
  \bibinfo{journal}{Phys. Rev. Lett.} \textbf{\bibinfo{volume}{26}},
  \bibinfo{pages}{331} (\bibinfo{year}{1971}),
  \urlprefix\url{http://link.aps.org/doi/10.1103/PhysRevLett.26.331}.

\bibitem[{\citenamefont{Yagi and Stein}(2016)}]{0264-9381-33-5-054001}
\bibinfo{author}{\bibfnamefont{K.}~\bibnamefont{Yagi}} \bibnamefont{and}
  \bibinfo{author}{\bibfnamefont{L.~C.} \bibnamefont{Stein}},
  \bibinfo{journal}{Classical and Quantum Gravity}
  \textbf{\bibinfo{volume}{33}}, \bibinfo{pages}{054001}
  (\bibinfo{year}{2016}),
  \urlprefix\url{http://stacks.iop.org/0264-9381/33/i=5/a=054001}.

\bibitem[{\citenamefont{Sotiriou}(2015)}]{0264-9381-32-21-214002}
\bibinfo{author}{\bibfnamefont{T.~P.} \bibnamefont{Sotiriou}},
  \bibinfo{journal}{Classical and Quantum Gravity}
  \textbf{\bibinfo{volume}{32}}, \bibinfo{pages}{214002}
  (\bibinfo{year}{2015}),
  \urlprefix\url{http://stacks.iop.org/0264-9381/32/i=21/a=214002}.

\bibitem[{\citenamefont{Dreyer et~al.}(2004)\citenamefont{Dreyer, Kelly,
  Krishnan, Finn, Garrison, and Lopez-Aleman}}]{0264-9381-21-4-003}
\bibinfo{author}{\bibfnamefont{O.}~\bibnamefont{Dreyer}},
  \bibinfo{author}{\bibfnamefont{B.}~\bibnamefont{Kelly}},
  \bibinfo{author}{\bibfnamefont{B.}~\bibnamefont{Krishnan}},
  \bibinfo{author}{\bibfnamefont{L.~S.} \bibnamefont{Finn}},
  \bibinfo{author}{\bibfnamefont{D.}~\bibnamefont{Garrison}}, \bibnamefont{and}
  \bibinfo{author}{\bibfnamefont{R.}~\bibnamefont{Lopez-Aleman}},
  \bibinfo{journal}{Classical and Quantum Gravity}
  \textbf{\bibinfo{volume}{21}}, \bibinfo{pages}{787} (\bibinfo{year}{2004}),
  \urlprefix\url{http://stacks.iop.org/0264-9381/21/i=4/a=003}.

\bibitem[{\citenamefont{Price and Pullin}(1994)}]{PhysRevLett.72.3297}
\bibinfo{author}{\bibfnamefont{R.~H.} \bibnamefont{Price}} \bibnamefont{and}
  \bibinfo{author}{\bibfnamefont{J.}~\bibnamefont{Pullin}},
  \bibinfo{journal}{Phys. Rev. Lett.} \textbf{\bibinfo{volume}{72}},
  \bibinfo{pages}{3297} (\bibinfo{year}{1994}),
  \urlprefix\url{http://link.aps.org/doi/10.1103/PhysRevLett.72.3297}.

\bibitem[{\citenamefont{Vishveshwara}(1970)}]{4783}
\bibinfo{author}{\bibfnamefont{C.~V.} \bibnamefont{Vishveshwara}},
  \bibinfo{journal}{Nature} \textbf{\bibinfo{volume}{227}},
  \bibinfo{pages}{936} (\bibinfo{year}{1970}).

\bibitem[{\citenamefont{{Teukolsky}}(1973)}]{1973ApJ...185..635T}
\bibinfo{author}{\bibfnamefont{S.~A.} \bibnamefont{{Teukolsky}}},
  \bibinfo{journal}{\apj} \textbf{\bibinfo{volume}{185}}, \bibinfo{pages}{635}
  (\bibinfo{year}{1973}).

\bibitem[{\citenamefont{S.~Chandrasekhar}(1975)}]{10.2307/78902}
\bibinfo{author}{\bibfnamefont{S.~D.} \bibnamefont{S.~Chandrasekhar}},
  \bibinfo{journal}{Proceedings of the Royal Society of London. Series A,
  Mathematical and Physical Sciences} \textbf{\bibinfo{volume}{344}},
  \bibinfo{pages}{441} (\bibinfo{year}{1975}), ISSN \bibinfo{issn}{00804630},
  \urlprefix\url{http://www.jstor.org/stable/78902}.

\bibitem[{\citenamefont{Echeverria}(1989)}]{PhysRevD.40.3194}
\bibinfo{author}{\bibfnamefont{F.}~\bibnamefont{Echeverria}},
  \bibinfo{journal}{Phys. Rev. D} \textbf{\bibinfo{volume}{40}},
  \bibinfo{pages}{3194} (\bibinfo{year}{1989}),
  \urlprefix\url{http://link.aps.org/doi/10.1103/PhysRevD.40.3194}.

\bibitem[{\citenamefont{Nollert}(1999)}]{Nollert:1999ji}
\bibinfo{author}{\bibfnamefont{H.-P.} \bibnamefont{Nollert}},
  \bibinfo{journal}{Class. Quant. Grav.} \textbf{\bibinfo{volume}{16}},
  \bibinfo{pages}{R159} (\bibinfo{year}{1999}).

\bibitem[{\citenamefont{Creighton}(1999)}]{PhysRevD.60.022001}
\bibinfo{author}{\bibfnamefont{J.~D.~E.} \bibnamefont{Creighton}},
  \bibinfo{journal}{Phys. Rev. D} \textbf{\bibinfo{volume}{60}},
  \bibinfo{pages}{022001} (\bibinfo{year}{1999}),
  \urlprefix\url{http://link.aps.org/doi/10.1103/PhysRevD.60.022001}.

\bibitem[{\citenamefont{Leaver}(1985)}]{Leaver285}
\bibinfo{author}{\bibfnamefont{E.~W.} \bibnamefont{Leaver}},
  \bibinfo{journal}{Proceedings of the Royal Society of London A: Mathematical,
  Physical and Engineering Sciences} \textbf{\bibinfo{volume}{402}},
  \bibinfo{pages}{285} (\bibinfo{year}{1985}), ISSN \bibinfo{issn}{0080-4630},
  \eprint{http://rspa.royalsocietypublishing.org/content/402/1823/285.full.pdf},
  \urlprefix\url{http://rspa.royalsocietypublishing.org/content/402/1823/285}.

\bibitem[{\citenamefont{{Miller} et~al.}(2015)\citenamefont{{Miller},
  {Barsotti}, {Vitale}, {Fritschel}, {Evans}, and
  {Sigg}}}]{2015PhRvD..91f2005M}
\bibinfo{author}{\bibfnamefont{J.}~\bibnamefont{{Miller}}},
  \bibinfo{author}{\bibfnamefont{L.}~\bibnamefont{{Barsotti}}},
  \bibinfo{author}{\bibfnamefont{S.}~\bibnamefont{{Vitale}}},
  \bibinfo{author}{\bibfnamefont{P.}~\bibnamefont{{Fritschel}}},
  \bibinfo{author}{\bibfnamefont{M.}~\bibnamefont{{Evans}}}, \bibnamefont{and}
  \bibinfo{author}{\bibfnamefont{D.}~\bibnamefont{{Sigg}}},
  \bibinfo{journal}{\prd} \textbf{\bibinfo{volume}{91}}, \bibinfo{eid}{062005}
  (\bibinfo{year}{2015}), \eprint{1410.5882}.

\bibitem[{\citenamefont{Hild et~al.}(2011)\citenamefont{Hild, Abernathy,
  Acernese, Amaro-Seoane, Andersson, Arun, Barone, Barr, Barsuglia, Beker
  et~al.}}]{0264-9381-28-9-094013}
\bibinfo{author}{\bibfnamefont{S.}~\bibnamefont{Hild}},
  \bibinfo{author}{\bibfnamefont{M.}~\bibnamefont{Abernathy}},
  \bibinfo{author}{\bibfnamefont{F.}~\bibnamefont{Acernese}},
  \bibinfo{author}{\bibfnamefont{P.}~\bibnamefont{Amaro-Seoane}},
  \bibinfo{author}{\bibfnamefont{N.}~\bibnamefont{Andersson}},
  \bibinfo{author}{\bibfnamefont{K.}~\bibnamefont{Arun}},
  \bibinfo{author}{\bibfnamefont{F.}~\bibnamefont{Barone}},
  \bibinfo{author}{\bibfnamefont{B.}~\bibnamefont{Barr}},
  \bibinfo{author}{\bibfnamefont{M.}~\bibnamefont{Barsuglia}},
  \bibinfo{author}{\bibfnamefont{M.}~\bibnamefont{Beker}},
  \bibnamefont{et~al.}, \bibinfo{journal}{Classical and Quantum Gravity}
  \textbf{\bibinfo{volume}{28}}, \bibinfo{pages}{094013}
  (\bibinfo{year}{2011}),
  \urlprefix\url{http://stacks.iop.org/0264-9381/28/i=9/a=094013}.

\bibitem[{\citenamefont{Evans et~al.}(2016)}]{Evans:2016dc}
\bibinfo{author}{\bibfnamefont{M.}~\bibnamefont{Evans}} \bibnamefont{et~al.},
  \bibinfo{journal}{www.dcc.ligo.org/P16000143}  (\bibinfo{year}{2016}).

\bibitem[{\citenamefont{Shahram and Milanfar}(2005)}]{1453789}
\bibinfo{author}{\bibfnamefont{M.}~\bibnamefont{Shahram}} \bibnamefont{and}
  \bibinfo{author}{\bibfnamefont{P.}~\bibnamefont{Milanfar}},
  \bibinfo{journal}{IEEE Transactions on Signal Processing}
  \textbf{\bibinfo{volume}{53}}, \bibinfo{pages}{2579} (\bibinfo{year}{2005}),
  ISSN \bibinfo{issn}{1053-587X}.

\bibitem[{\citenamefont{Milanfar and Shakouri}(2002)}]{1038162}
\bibinfo{author}{\bibfnamefont{P.}~\bibnamefont{Milanfar}} \bibnamefont{and}
  \bibinfo{author}{\bibfnamefont{A.}~\bibnamefont{Shakouri}}, in
  \emph{\bibinfo{booktitle}{Image Processing. 2002. Proceedings. 2002
  International Conference on}} (\bibinfo{year}{2002}),
  vol.~\bibinfo{volume}{1}, pp. \bibinfo{pages}{I--864--I--867 vol.1}, ISSN
  \bibinfo{issn}{1522-4880}.

\bibitem[{\citenamefont{The LIGO
  Scientific~Collaboration}(2016)}]{LIGO:2016Rates}
\bibinfo{author}{\bibfnamefont{T.~V.~C.} \bibnamefont{The LIGO
  Scientific~Collaboration}}, \bibinfo{journal}{www.dcc.ligo.org/P1600088}
  (\bibinfo{year}{2016}).

\bibitem[{\citenamefont{{LIGO Scientific Collaboration}
  et~al.}(2015)\citenamefont{{LIGO Scientific Collaboration}, {Aasi}, {Abbott},
  {Abbott}, {Abbott}, {Abernathy}, {Ackley}, {Adams}, {Adams}, {Addesso}
  et~al.}}]{2015CQGra..32g4001L}
\bibinfo{author}{\bibnamefont{{LIGO Scientific Collaboration}}},
  \bibinfo{author}{\bibfnamefont{J.}~\bibnamefont{{Aasi}}},
  \bibinfo{author}{\bibfnamefont{B.~P.} \bibnamefont{{Abbott}}},
  \bibinfo{author}{\bibfnamefont{R.}~\bibnamefont{{Abbott}}},
  \bibinfo{author}{\bibfnamefont{T.}~\bibnamefont{{Abbott}}},
  \bibinfo{author}{\bibfnamefont{M.~R.} \bibnamefont{{Abernathy}}},
  \bibinfo{author}{\bibfnamefont{K.}~\bibnamefont{{Ackley}}},
  \bibinfo{author}{\bibfnamefont{C.}~\bibnamefont{{Adams}}},
  \bibinfo{author}{\bibfnamefont{T.}~\bibnamefont{{Adams}}},
  \bibinfo{author}{\bibfnamefont{P.}~\bibnamefont{{Addesso}}},
  \bibnamefont{et~al.}, \bibinfo{journal}{Classical and Quantum Gravity}
  \textbf{\bibinfo{volume}{32}}, \bibinfo{eid}{074001} (\bibinfo{year}{2015}),
  \eprint{1411.4547}.

\bibitem[{\citenamefont{{Berti} et~al.}(2016)\citenamefont{{Berti}, {Sesana},
  {Barausse}, {Cardoso}, and {Belczynski}}}]{2016arXiv160509286B}
\bibinfo{author}{\bibfnamefont{E.}~\bibnamefont{{Berti}}},
  \bibinfo{author}{\bibfnamefont{A.}~\bibnamefont{{Sesana}}},
  \bibinfo{author}{\bibfnamefont{E.}~\bibnamefont{{Barausse}}},
  \bibinfo{author}{\bibfnamefont{V.}~\bibnamefont{{Cardoso}}},
  \bibnamefont{and}
  \bibinfo{author}{\bibfnamefont{K.}~\bibnamefont{{Belczynski}}},
  \bibinfo{journal}{ArXiv e-prints}  (\bibinfo{year}{2016}),
  \eprint{1605.09286}.

\bibitem[{\citenamefont{Van~Trees}(2004)}]{van2004detection}
\bibinfo{author}{\bibfnamefont{H.~L.} \bibnamefont{Van~Trees}},
  \emph{\bibinfo{title}{Detection, estimation, and modulation theory}}
  (\bibinfo{publisher}{John Wiley \& Sons}, \bibinfo{year}{2004}).

\bibitem[{\citenamefont{{Beyer}}(1999)}]{1999CMaPh.204..397B}
\bibinfo{author}{\bibfnamefont{H.~R.} \bibnamefont{{Beyer}}},
  \bibinfo{journal}{Communications in Mathematical Physics}
  \textbf{\bibinfo{volume}{204}}, \bibinfo{pages}{397} (\bibinfo{year}{1999}),
  \eprint{gr-qc/9803034}.

\bibitem[{\citenamefont{Kokkotas and Schmidt}(1999)}]{lrr-1999-2}
\bibinfo{author}{\bibfnamefont{K.~D.} \bibnamefont{Kokkotas}} \bibnamefont{and}
  \bibinfo{author}{\bibfnamefont{B.~G.} \bibnamefont{Schmidt}},
  \bibinfo{journal}{Living Reviews in Relativity} \textbf{\bibinfo{volume}{2}}
  (\bibinfo{year}{1999}),
  \urlprefix\url{http://www.livingreviews.org/lrr-1999-2}.

\bibitem[{\citenamefont{{Ruiz} et~al.}(2008)\citenamefont{{Ruiz}, {Alcubierre},
  {N{\'u}{\~n}ez}, and {Takahashi}}}]{2008GReGr..40.1705R}
\bibinfo{author}{\bibfnamefont{M.}~\bibnamefont{{Ruiz}}},
  \bibinfo{author}{\bibfnamefont{M.}~\bibnamefont{{Alcubierre}}},
  \bibinfo{author}{\bibfnamefont{D.}~\bibnamefont{{N{\'u}{\~n}ez}}},
  \bibnamefont{and}
  \bibinfo{author}{\bibfnamefont{R.}~\bibnamefont{{Takahashi}}},
  \bibinfo{journal}{General Relativity and Gravitation}
  \textbf{\bibinfo{volume}{40}}, \bibinfo{pages}{1705} (\bibinfo{year}{2008}),
  \eprint{0707.4654}.

\bibitem[{\citenamefont{{Berti} and {Cardoso}}(2006)}]{2006PhRvD..74j4020B}
\bibinfo{author}{\bibfnamefont{E.}~\bibnamefont{{Berti}}} \bibnamefont{and}
  \bibinfo{author}{\bibfnamefont{V.}~\bibnamefont{{Cardoso}}},
  \bibinfo{journal}{\prd} \textbf{\bibinfo{volume}{74}}, \bibinfo{eid}{104020}
  (\bibinfo{year}{2006}), \eprint{gr-qc/0605118}.

\bibitem[{\citenamefont{{Berti} and {Klein}}(2014)}]{2014PhRvD..90f4012B}
\bibinfo{author}{\bibfnamefont{E.}~\bibnamefont{{Berti}}} \bibnamefont{and}
  \bibinfo{author}{\bibfnamefont{A.}~\bibnamefont{{Klein}}},
  \bibinfo{journal}{\prd} \textbf{\bibinfo{volume}{90}}, \bibinfo{eid}{064012}
  (\bibinfo{year}{2014}), \eprint{1408.1860}.

\bibitem[{\citenamefont{Berti et~al.}(2006)\citenamefont{Berti, Cardoso, and
  Will}}]{PhysRevD.73.064030}
\bibinfo{author}{\bibfnamefont{E.}~\bibnamefont{Berti}},
  \bibinfo{author}{\bibfnamefont{V.}~\bibnamefont{Cardoso}}, \bibnamefont{and}
  \bibinfo{author}{\bibfnamefont{C.~M.} \bibnamefont{Will}},
  \bibinfo{journal}{Phys. Rev. D} \textbf{\bibinfo{volume}{73}},
  \bibinfo{pages}{064030} (\bibinfo{year}{2006}),
  \urlprefix\url{http://link.aps.org/doi/10.1103/PhysRevD.73.064030}.

\bibitem[{\citenamefont{{London} et~al.}(2014)\citenamefont{{London},
  {Shoemaker}, and {Healy}}}]{2014PhRvD..90l4032L}
\bibinfo{author}{\bibfnamefont{L.}~\bibnamefont{{London}}},
  \bibinfo{author}{\bibfnamefont{D.}~\bibnamefont{{Shoemaker}}},
  \bibnamefont{and} \bibinfo{author}{\bibfnamefont{J.}~\bibnamefont{{Healy}}},
  \bibinfo{journal}{\prd} \textbf{\bibinfo{volume}{90}}, \bibinfo{eid}{124032}
  (\bibinfo{year}{2014}), \eprint{1404.3197}.

\bibitem[{\citenamefont{Cutler and Flanagan}(1994)}]{PhysRevD.49.2658}
\bibinfo{author}{\bibfnamefont{C.}~\bibnamefont{Cutler}} \bibnamefont{and}
  \bibinfo{author}{\bibfnamefont{E.~E.} \bibnamefont{Flanagan}},
  \bibinfo{journal}{Phys. Rev. D} \textbf{\bibinfo{volume}{49}},
  \bibinfo{pages}{2658} (\bibinfo{year}{1994}),
  \urlprefix\url{http://link.aps.org/doi/10.1103/PhysRevD.49.2658}.

\bibitem[{\citenamefont{Finn}(1992)}]{PhysRevD.46.5236}
\bibinfo{author}{\bibfnamefont{L.~S.} \bibnamefont{Finn}},
  \bibinfo{journal}{Phys. Rev. D} \textbf{\bibinfo{volume}{46}},
  \bibinfo{pages}{5236} (\bibinfo{year}{1992}),
  \urlprefix\url{http://link.aps.org/doi/10.1103/PhysRevD.46.5236}.

\bibitem[{\citenamefont{{Dreyer} et~al.}(2004)\citenamefont{{Dreyer}, {Kelly},
  {Krishnan}, {Finn}, {Garrison}, and {Lopez-Aleman}}}]{2004CQGra..21..787D}
\bibinfo{author}{\bibfnamefont{O.}~\bibnamefont{{Dreyer}}},
  \bibinfo{author}{\bibfnamefont{B.}~\bibnamefont{{Kelly}}},
  \bibinfo{author}{\bibfnamefont{B.}~\bibnamefont{{Krishnan}}},
  \bibinfo{author}{\bibfnamefont{L.~S.} \bibnamefont{{Finn}}},
  \bibinfo{author}{\bibfnamefont{D.}~\bibnamefont{{Garrison}}},
  \bibnamefont{and}
  \bibinfo{author}{\bibfnamefont{R.}~\bibnamefont{{Lopez-Aleman}}},
  \bibinfo{journal}{Classical and Quantum Gravity}
  \textbf{\bibinfo{volume}{21}}, \bibinfo{pages}{787} (\bibinfo{year}{2004}),
  \eprint{gr-qc/0309007}.

\bibitem[{\citenamefont{{Kamaretsos} et~al.}(2012)\citenamefont{{Kamaretsos},
  {Hannam}, and {Sathyaprakash}}}]{2012PhRvL.109n1102K}
\bibinfo{author}{\bibfnamefont{I.}~\bibnamefont{{Kamaretsos}}},
  \bibinfo{author}{\bibfnamefont{M.}~\bibnamefont{{Hannam}}}, \bibnamefont{and}
  \bibinfo{author}{\bibfnamefont{B.~S.} \bibnamefont{{Sathyaprakash}}},
  \bibinfo{journal}{Physical Review Letters} \textbf{\bibinfo{volume}{109}},
  \bibinfo{eid}{141102} (\bibinfo{year}{2012}), \eprint{1207.0399}.

\bibitem[{\citenamefont{{Berti} et~al.}(2007)\citenamefont{{Berti}, {Cardoso},
  {Gonzalez}, {Sperhake}, {Hannam}, {Husa}, and
  {Br{\"u}gmann}}}]{2007PhRvD..76f4034B}
\bibinfo{author}{\bibfnamefont{E.}~\bibnamefont{{Berti}}},
  \bibinfo{author}{\bibfnamefont{V.}~\bibnamefont{{Cardoso}}},
  \bibinfo{author}{\bibfnamefont{J.~A.} \bibnamefont{{Gonzalez}}},
  \bibinfo{author}{\bibfnamefont{U.}~\bibnamefont{{Sperhake}}},
  \bibinfo{author}{\bibfnamefont{M.}~\bibnamefont{{Hannam}}},
  \bibinfo{author}{\bibfnamefont{S.}~\bibnamefont{{Husa}}}, \bibnamefont{and}
  \bibinfo{author}{\bibfnamefont{B.}~\bibnamefont{{Br{\"u}gmann}}},
  \bibinfo{journal}{\prd} \textbf{\bibinfo{volume}{76}}, \bibinfo{eid}{064034}
  (\bibinfo{year}{2007}), \eprint{gr-qc/0703053}.

\bibitem[{\citenamefont{Gossan et~al.}(2012)\citenamefont{Gossan, Veitch, and
  Sathyaprakash}}]{PhysRevD.85.124056}
\bibinfo{author}{\bibfnamefont{S.}~\bibnamefont{Gossan}},
  \bibinfo{author}{\bibfnamefont{J.}~\bibnamefont{Veitch}}, \bibnamefont{and}
  \bibinfo{author}{\bibfnamefont{B.~S.} \bibnamefont{Sathyaprakash}},
  \bibinfo{journal}{Phys. Rev. D} \textbf{\bibinfo{volume}{85}},
  \bibinfo{pages}{124056} (\bibinfo{year}{2012}),
  \urlprefix\url{http://link.aps.org/doi/10.1103/PhysRevD.85.124056}.

\bibitem[{\citenamefont{{London}}(2016)}]{Lionel:comm}
\bibinfo{author}{\bibfnamefont{L.}~\bibnamefont{{London}}},
  \bibinfo{journal}{Private Communication}  (\bibinfo{year}{2016}).

\end{thebibliography}

\end{document}